\documentclass{aa}  

\usepackage{graphicx}
\usepackage{txfonts}
\usepackage{xspace}
\usepackage{upgreek}

\newcommand{\F}{\textit{Fermi}\xspace}
\newcommand{\highlight}[1]{{#1}\xspace}

\begin{document} 


\title{Disentangling multiple high-energy emission components in the Vela~X pulsar wind nebula with the \F Large Area Telescope}
\titlerunning{Vela~X multiple components with \F~LAT}
\author{
L.~Tibaldo$^{(1)}$ \and 
R.~Zanin$^{(2)}$ \and
G.~Faggioli$^{(3)}$ \and
J.~Ballet$^{(4)}$ \and  
M.-H.~Grondin$^{(5)}$ \and 
J.~A.~Hinton$^{(2)}$ \and 
M.~Lemoine-Goumard$^{(5)}$
}
\authorrunning{Tibaldo et al.}

\institute{
\inst{1}~IRAP, Universit\'e de Toulouse, CNRS, UPS, CNES, F-31028 Toulouse, France\\ 
\inst{2}~Max-Planck-Institut f\"ur Kernphysik, D-69029 Heidelberg, Germany\\
\inst{3}~Dipartimento di Fisica, Universit\`a degli Studi di Torino, I-10125 Torino, Italy\\ 
\inst{4}~Laboratoire AIM, CEA-IRFU/CNRS/Universit\'e Paris Diderot, Service d'Astrophysique, CEA Saclay, F-91191 Gif sur Yvette, France\\ 
\inst{5}~Centre d'\'Etudes Nucl\'eaires de Bordeaux Gradignan, IN2P3/CNRS, Universit\'e Bordeaux 1, BP120, F-33175 Gradignan Cedex, France\\  
\email{luigi.tibaldo@irap.omp.eu} \\
}
  \abstract
   {Vela~X is a pulsar wind nebula in which two relativistic particle populations with distinct spatial and spectral distributions dominate the emission at different wavelengths. An extended $2\degr \times 3\degr$ nebula is seen in radio and GeV gamma rays. An elongated cocoon prevails in X-rays and TeV gamma rays.}
   {We use $\sim$9.5~years of data from the \F~Large Area Telescope (LAT) to disentangle gamma-ray emission from the two components in the energy range from 10~GeV to 2~TeV, bridging the gap between previous measurements at GeV and TeV energies.}
   {We determine the morphology of emission associated to Vela~X separately at energies $<100$~GeV and $>100$~GeV, and compare it to the morphology seen at other wavelengths. Then, we derive the spectral energy distribution of the two gamma-ray components over the full energy range.}
   {The best overall fit to the LAT data is provided by the combination of the two components derived at energies $<100$~GeV and $>100$~GeV. The first component has a soft spectrum, spectral index $2.19\pm0.16^{+0.05}_{-0.22}$, and extends over a region of radius $1\fdg36\pm0\fdg04$, consistent with the size of the radio nebula. The second component has a harder spectrum, spectral index $0.9\pm0.3^{+0.3}_{-0.1}$, and is concentrated over an area of radius $0\fdg63\pm0\fdg03$, coincident with the X-ray cocoon that had already been established to account for the bulk of the emission at TeV energies.}
   {The spectrum measured for the low-energy component corroborates previous evidence for a roll-over of the electron spectrum in the extended radio nebula at energies of a few tens of GeV possibly due to diffusive escape. The high-energy component has a very hard spectrum: if the emission is produced by electrons with a power-law spectrum the electrons must be uncooled, and there is a hint that their spectrum may be harder than predictions by standard models of Fermi acceleration at relativistic shocks.}


   \maketitle
%

\section{Introduction}

Pulsars generate powerful winds that form nebulae filled by magnetized relativistic plasma, \highlight{which}, in turn, produce nonthermal radiation from radio to gamma rays. Pulsar Wind Nebulae \citep[PWNe, e.g.,][]{pwnreview13} are a prime observational target to understand particle acceleration in relativistic plasmas and are plausibly an important source of \highlight{high-energy} cosmic ray electrons and positrons. 

Vela~X is the PWN formed by the Vela pulsar, \object{PSR~J0835$-$4510}. Due to its modest distance from the Sun of only $287^{+19}_{-17}$~pc \citep[][from parallax measurement of the pulsar]{dodson2003}, Vela X is spatially resolved at many wavelengths. Two main components appear to dominate depending on the wavelength \citep[see, e.g.,][and references therein]{hinton2011}. From radio to microwaves, Vela~X appears as an extended $2\degr \times 3\degr$ nebula with a butterfly-like morphology. \highlight{also consistent with GeV gamma-ray observations}. The pulsar lies at the position of the butterfly head, and we observe a series of filaments departing from it along the butterfly body, forming an elongated $1\degr$ structure dubbed the cocoon, that dominates emission in X-rays and TeV gamma rays. 

The conventional interpretation originally advocated by \citet{dejager2008} is that two populations of accelerated electrons with different spectra coexist in Vela~X. \citet{hinton2011} proposed that the extended radio nebula shelters an older electron population accelerated in the first phases of the pulsar and supernova remnant (SNR) life, and that the highest-energy particles have by now left this region due to diffusive escape. On the other hand, the cocoon would be filled by an electron population accelerated more recently and not affected yet by escape.

This hypothesis is motivated by the prediction of multiple particle populations in middle-aged PWNe evolving in SNRs by \citet{gelfand2009}, and by the hydrodynamical simulations of \citet{blondin2001}, who suggested that the Vela~X cocoon is a recent structure formed $\sim$20~kyr after the explosion of the supernova and the birth of the pulsar, due to the SNR reverse shock crushing the PWN. This scenario also justifies the \highlight{morphology of the radio nebula and the off-center position of the pulsar}, due to the asymmetric interaction with the SNR. The timescale suggested by the hydrodynamical simulations is in agreement with the age of the pulsar of 20-30~kyr inferred from its spindown history \citep{lyne1996}.

The gamma-ray spectrum and morphology of Vela~X have been studied in detail so far only at energies $< 100$~GeV using the Large Area Telescope (LAT) aboard the \textit{Fermi Gamma-ray Space Telescope} \citep{fermivelax2010,grondin2013}, and at energies $>550$~GeV using the H.E.S.S. array of atmospheric Cherenkov telescopes \citep{hessvelax2006,hessvelax2012}. \citet{2FHL,3FHL} have shown that the most recent \F~LAT dataset based on the \textit{Pass~8} event-level analysis \citep{atwood2013} in fact shows significant emission associated with Vela~X at all energies from 10~GeV up to 2~TeV.

In this article we present a new analysis of \F~LAT data $> 10$~GeV, and use it to disentangle the spectra of the different morphological components in Vela~X. Thus, we bridge the gap between previous measurements, and we probe for the first time the spectral distribution of the highest-energy particles in the extended nebula and of the lowest-energy particles in the cocoon.

\section{Observations, Analysis, and Results}

\subsection{Dataset and Analysis Generalities}

The \F~LAT is an imaging pair-tracking telescope that detects gamma rays from 30~MeV to $>1$~TeV \citep{atwood2009}. We use all observations performed using the LAT from MET\footnote{\F Mission Elapsed Time, i.e., seconds since the reference time of January 1, 2001, at midnight in the Coordinated Universal Time (UTC) system.} 239610747~s to 541486510~s, i.e., over $\sim$9.5~years of operations.

We use all events belonging to the \textit{Pass 8 Source} class with reconstructed direction within 105\degr\ from the local zenith (to eliminate emission from the Earth's atmosphere) and reconstructed energies $>10$~GeV and $<2$~TeV. The lower energy limit of 10~GeV minimizes the contamination from the bright Vela pulsar, because the spectrum of the pulsar has a cutoff at 3.0~GeV \citep{2PC}, and also the LAT Point Spread Function\footnote{\url{http://www.slac.stanford.edu/exp/glast/groups/canda/lat_Performance.htm}.} (PSF) 68\% containment radius attains values $<0\fdg15$, much smaller than the size of either component of the PWN. Furthermore, the energy threshold of 10 GeV reduces confusion with the bright interstellar emission from the Milky Way, \highlight{which} has a spectrum softer than Vela X.

We perform a binned maximum likelihood analysis using Poisson statistics. The analysis region is a $5\degr \times 5\degr$ square in Celestial coordinates (J2000 equinox) with Aitoff projection centered at the position of Vela pulsar. We bin the events over a 0\fdg1 grid spatially, and using 8 bins per decade in energy. For the analysis we employ the \F~LAT Science Tools 11-05-03 and the Python package Fermipy \citep{wood2017} version 0.15.1. We use the LAT Instrument Response Functions (IRFs) \texttt{P8R2\_SOURCE\_V6}.

The starting model to be fit to the observations is constructed combining all sources in the most recent LAT high-energy catalogs\highlight{: 2FHL \citep{2FHL} and 3FHL \citep{3FHL}. \highlight{For sources present in both catalogs we take the model from the more recent 3FHL, also consistent with our energy range.} We also include} a model for interstellar emission produced by cosmic-ray interactions with gas and radiation fields in the Milky Way \citep{IEM} and a spectral model for the isotropic background that accounts for extragalactic diffuse gamma-ray emission and the residual background from cosmic-ray interactions in the LAT misclassified as gamma rays\footnote{The two models are \texttt{gll\_iem\_v06.fits} and \texttt{iso\_P8R2\_SOURCE\_V6\_v06.txt}, respectively, see \url{https://fermi.gsfc.nasa.gov/ssc/data/access/lat/BackgroundModels.html}.}. We fix the normalization of the isotropic background to~1, and leave free the normalization of the interstellar emission model, which has a higher number of predicted event counts for our region.

\subsection{Morphological Analysis in the Low- and High-Energy Domains}\label{sec:ebands}

We start by subdividing the whole energy range into a low-energy (LE) range from 10~GeV to 100~GeV, and a high-energy (HE) range from 100~GeV to 2~TeV. We choose 100~GeV based on the extrapolation of the spectra from \citet{hessvelax2012,grondin2013} and the spectral models in \citet{hinton2011} because we expect the extended nebula/cocoon to be dominant below/above this energy. We have verified that the results presented in the article do not depend critically on the exact value of the energy threshold around 100~GeV.

For each energy range we separately derive from the data the morphology of the emission associated with Vela~X. \highlight{First of all, we re-evaluate the significance of all sources in the input catalogs, as quantified through the likelihood ratio test. We compute the Test Statistic $\mathrm{TS} = 2 \times \Delta \log \mathcal{L} = 2 \times (\log \mathcal{L}_\mathrm{test} - \log \mathcal{L}_0)$, where $\mathcal{L}_\mathrm{test}$ and $\mathcal{L}_0$ are the maximum likelihood values of the test hypothesis and of the null hypothesis, respectively, i.e., for this application, the model including or not including the source.} Sources with $\mathrm{TS}<2$ are removed from the model, and sources with $\mathrm{TS}<9$ have their spectral parameters fixed to the catalog values. A point source at the position of the Vela pulsar is always left in the model. \highlight{Following the 3FHL catalog \citep{3FHL}, the spectrum of the Vela pulsar is modeled as a log-parabola. The spectral parameters of the Vela pulsar are left free in the LE range, where $\mathrm{TS}>9$, and fixed to the determination in the catalog in the HE range, where $\mathrm{TS}<9$.} The catalog model of Vela X is removed.

We then apply { Fermipy's} iterative point-like \highlight{source-finding} algorithm { \citep[][Section~4.5]{wood2017}} that uses peak finding on a TS map to define new source candidates. Peaks with a minimum TS of 9 and minimum separation of 0\fdg5 from other sources in the model are considered as candidates, and their position is determined by fitting a 2D parabola to the log-likelihood surface around the peak under the hypothesis of a power-law spectrum. A maximum of 10 sources per iteration are added, and the process is iterated until no additional candidates are found or up to a maximum of 5 iterations. We find a total of 16 source candidates in the LE domain and 4 source candidates in the HE domain, respectively.

Then, we test the hypothesis that newly found point-like source candidates within Vela~X are associated with emission from the extended PWN. They are identified by their positions being in a region with brightness temperature $> 2.5$~K in the 330~MHz VLA image in \citet{frail1997}. Those sources, 11 in the LE range and 3 in the HE range, are removed from the model, and replaced by an extended disk centered at the TS-weighted barycenter of the point sources positions. The extension and center of the disk are then simultaneously fit to the data through a multidimensional likelihood profile scan. This procedure makes it possible to derive the best-fit morphology from the data without being biased by the extended models of Vela~X from previous analyses of LAT data. In the LE range the 11 sources are distributed uniformly across the area of the final best-fit disk, while in the HE range the 3 sources are aligned along the direction of the X-ray cocoon. The spectral indices of the point-like sources have very large uncertainties due to the limited statistics, so that their values within each energy range are formally compatible with each other. { The remaining source candidates found outside Vela~X are listed in Table~\ref{tab:candidate-sources}.}

The results from the morphological analyses are reported in Table~\ref{tab:morphology}, \highlight{where the differences in likelihood and number of degrees of freedom are used below}.
\begin{table*}
\caption{\label{tab:morphology}Results from the Morphological Analysis of Vela~X in the Low- and High-Energy Domains.}
\centering
\begin{tabular}{lcc}
\hline\hline
										& LE ($<100$~GeV)				& HE ($> 100$~GeV)\\
\hline                                        
RA (J2000)								& $128\fdg65\pm0\fdg11$		& $128\fdg06\pm0\fdg05$\\
Dec (J2000)								& $-45\fdg47\pm0\fdg06$		& $-45\fdg84\pm0\fdg05$\\
$l$										& $263\fdg71\pm0\fdg08$		& $263\fdg76\pm0\fdg06$\\
$b$										& $-3\fdg07\pm0\fdg09$		& $-3\fdg62\pm0\fdg04$\\
Center error\tablefootmark{a} ($1\sigma$)				& $0\fdg13$					& $0\fdg07$\\
radius									& $1\fdg36\pm0\fdg04$		& $0\fdg63\pm0\fdg03$\\
\hline
$\Delta \log \mathcal{L}$ (d.o.f.), disk/no source& 89.9 (5)						& 30.5 (5)\\
$\Delta \log \mathcal{L}$ (d.o.f.), disk/point source & 78.0 (1) & 15.2 (1)\\
$\Delta \log \mathcal{L}$ (d.o.f.), disk/multiple point sources\tablefootmark{b} & -23.2 (-39) & 0.18 (-7)\\
$\Delta \log \mathcal{L}$ (d.o.f.), disk/template\tablefootmark{c} & -2.1 (3) & 11.7 (3)\\
\hline
\end{tabular}
\tablefoot{We report the parameters of the best-fit disks (top), and the \highlight{differences} in maximum-likelihood logarithm ($\log \mathcal{L}$) and number of degrees of freedom (d.o.f) between the disk models and the hypothesis that there is no emission associated with Vela~X, and other models considered in the analysis (bottom).\\
\tablefoottext{a}{Statistical uncertainty in the disk center position.}
\tablefoottext{b}{The multiple point source model has 11 sources associated with Vela X in the LE domain, and 3 in the HE domain.}
\tablefoottext{c}{\highlight{We compare the best-fit disks with the radio template in the LE range, and the H.E.S.S. template in the HE range, respectively. See section \ref{sec:ebands} for details.}}
}
\end{table*}
Extended emission spatially associated with Vela~X is significantly detected in both energy ranges. This is assessed quantitatively using the likelihood ratio test, i.e., through the values of $\mathrm{TS} = 2 \times \Delta \log \mathcal{L}$ for the disk hypothesis compared to the hypothesis of no emission associated with Vela X, and for the disk hypothesis compared to the point source hypothesis. Although we do not meet all criteria necessary to apply the likelihood ratio test \citep[in both cases the null hypothesis lies on the border of the parameter space of the test hypothesis, see][]{protassov2002}, based on the Monte~Carlo studies in \citet{mattox1996} and \citet{lande2012} we assume that in the null hypothesis TS is distributed as $\chi^2_n/2$, where the number of degrees of freedom is 5 (1) for the disk hypothesis compared to the hypothesis of no emission associated with Vela X (for the disk hypothesis compared to the point source hypothesis), i.e., the difference in number of degrees of freedom between test and null hypothesis. Following this approach the significance of the detection of the extended disks is estimated as $13\sigma$ and $7.3\sigma$ in the LE and HE range, respectively. Additionally, the extended disk hypothesis is preferred over the hypothesis of a single point source at $12.5\sigma$ and $5.5\sigma$ confidence level in the LE and HE range, respectively.

We cannot use the likelihood ratio test to compare the hypothesis of an extended disk to that of multiple point sources, because the two models are not nested \citep[e.g.,][]{protassov2002}. However, we can use the Akaike information criterion \citep[AIC,][]{AIC}. We calculate $\mathrm{AIC} = 2\times (\Delta\, \mathrm{d.o.f} - \Delta \log \mathcal{L})$, which results to be $-31.6$ ($-13.6$) in the LE (HE) domain. The negative values indicate that the disk model is preferred over the model with multiple point sources in both energy ranges. { Note that, however, the AIC does not enable us to quantify the significance of the preference of one model over the other.}

   \begin{figure*}
   \centering
  \includegraphics[width=1\textwidth]{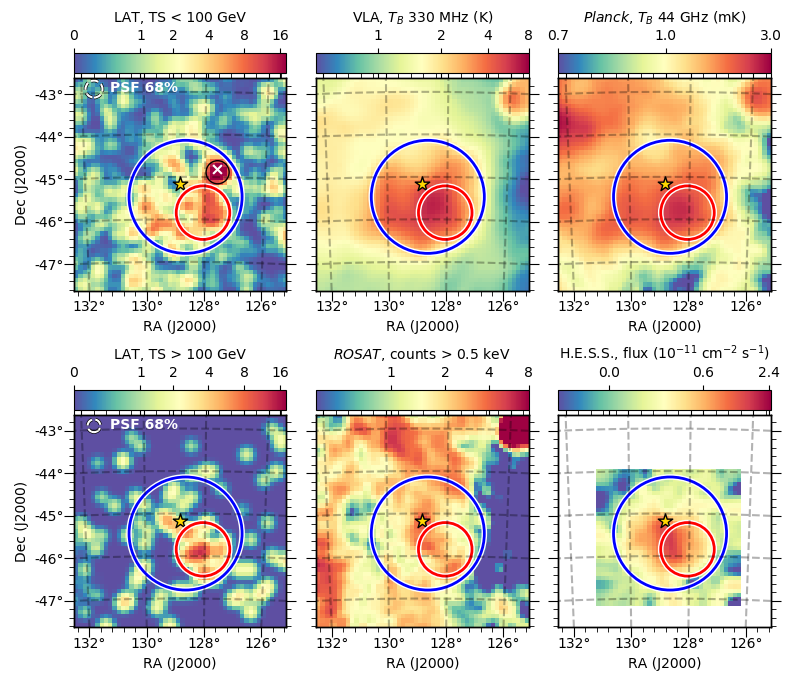}
   \caption{Multiwavelength maps of the analysis region. The LAT TS maps for energies below and above 100 GeV are derived as described in Section~\ref{sec:ebands}. The VLA map at 330 MHz is reproduced from \citet{frail1997}, the H.E.S.S. map for energies $>750$~GeV from \citet{hessvelax2012}. We also show maps at 44~GHz from the \textit{Planck} survey \citep{planck2015}, and at X-ray energies $>0.5$~keV from the \textit{ROSAT} survey \citep{RASS}. The blue and red circles show the best-fit disks that account for Vela~X in LAT data at energies $<100$~GeV and $>100$~GeV, respectively (Section~\ref{sec:ebands}). The star shows the position of the Vela pulsar. The black circle in the top left plot corresponds to the source FGES~J0830.3$-$4453 \citep{FGES}, while the white cross shows the position of the source PS~J0830.4$-$4449 found in our analysis. Note that for the LAT maps all sources other than Vela~X and FGES~J0830.3$-$4453/PS~J0830.4$-$4449, including the Vela pulsar, were accounted for in the background model. All maps are reprojected on the same grid as the gamma-ray images. Furthermore, they are smoothed for display to a common resolution of $0\fdg15$. In the top left corner of the LAT maps we show the effective PSF~68\% containment circle in the corresponding energy range for a power-law spectral distribution with index~2. 
   }
              \label{fig:maps}%
    \end{figure*}

The best-fit disk center and extension differ between the LE and HE domains (Table~\ref{tab:morphology}). This is also illustrated in Figure~\ref{fig:maps}. In the figure we show the TS maps in the two energy ranges obtained from the LAT data based on a model that does not include extended emission from Vela~X (the disk is removed from the best-fit model). We calculate on the map grid the likelihood of a model with an additional point source with a power-law spectrum with spectral index of~2. Then, we overlay to the TS map the contour of the best-fit disks. We also compare the LAT TS maps and best-fit disks with multiwavelength maps of the region. In the LE range the emission overlaps the whole region of the extended nebula as seen in radio/microwaves, \highlight{consistent} with previous results \citep{fermivelax2010,grondin2013}. At higher energies the emission becomes more concentrated and is coincident with the X-ray cocoon. The HE emission also overlaps the bulk of the emission seen at higher energies with H.E.S.S. \citep{hessvelax2006}. The extension from our analysis, 68\% containment radius of $0\fdg52\pm0\fdg02$, is slightly larger than what was measured with H.E.S.S., $0\fdg43\pm0\fdg02$. There is also a 0\fdg6 shift of the emission centroid.

We further assess the correlation with emission at other wavelengths by fitting templates to the LAT data:
\begin{itemize}
\item for the LE component we use the radio map from \citet[][Figure~\ref{fig:maps}]{frail1997}, on a $4\degr \times 4\degr$ region centered at R.A.~$=128\fdg6$, Dec~$=-45\fdg7$; a priori we do not expect an exact proportionality between gamma-ray intensities and radio intensities, because radio emission is produced by synchrotron emission due to magnetic fields, while gamma-ray emission results from inverse-Compton (IC) scattering of low-energy photons;
\item for the HE component we use the gamma-ray map from \citet{hessvelax2012}; since the high-energy PSF of the LAT is comparable to that of H.E.S.S., we apply a PSF deconvolution using the Richardson-Lucy algorithm \citep{lucy1974} as implemented in the \texttt{scikit-image} Python package \citep{scikit-image2014}; furthermore we use for the template only the region of 0\fdg8 radius where the significance of the gamma-ray emission detected with H.E.S.S. is the highest \citep{hessvelax2012}.
\end{itemize}

In the LE domain the likelihoods of the fit with the disk and the radio template are very similar, confirming that there is a very good correlation between gamma-ray and radio emission, with the radio template preferred based on the Akaike information criterion (AIC~$=-10.2$). In the HE domain the disk provides a better fit to the LAT data than the H.E.S.S. template, AIC~$=-17.4$, which is consistent with the shift already noted between emission measured by the LAT and H.E.S.S. In the HE domain the single disk is preferred even to the combination of radio and H.E.S.S. templates, with AIC~$=-6$. However, counting statistics for the LAT in the HE domain are too low to draw robust conclusions on this point, as well as with respect to the fainter TeV emission beyond the cocoon over an area with radius of 1\fdg2 detected by H.E.S.S. \citep{hessvelax2012}. 

In the LE TS map (Figure~\ref{fig:maps}) the highest peak corresponds to the object FGES~J0830.3$-$4453 from the LAT catalog of extended Galactic sources detected at energies $>10$~GeV \citep{FGES}. In \citet{FGES} the bulk of emission from Vela~X is accounted by the source FGES~J0832.0$-$4549, which has an extension radius of $0\fdg71\pm0\fdg05$, smaller than the best-fit extension of the disk that we obtained in the LE range, $1\fdg36\pm0\fdg04$. Our pointlike source-finding procedure had identified a candidate coincident with FGES~J0830.3$-$4453. This source candidate is one of the 11 that we have removed from the model owing to its overlap with radio emission from Vela~X. We have shown that our best-fit disk is preferred to the combination of the 11 point sources based on the AIC.

We further test the significance of an individual source coincident with FGES~J0830.3$-$4453 on top of the radio template, which provides the best representation of the large-scale emission from Vela~X in \highlight{the} LE range. Using a multidimensional likelihood profile scan we first localize a pointlike source starting from the previously-found seed, and then we simultaneously fit the center and radius of an extended disk source to the data. The TS value for the pointlike source hypothesis compared to the hypothesis of no individual source on top of the radio template is 29.4, therefore an individual source is detected with a significance of $4.5\sigma$. The TS value for the hypothesis of a disk source compared to the pointlike hypothesis is 7.1, therefore the source is not significantly extended ($2.7\sigma$). The pointlike source, hereafter referred to as PS~J0830.4$-$4449, is localized at R.A.~$=127\fdg61\pm 0\fdg03$, Dec~$=-44\fdg82\pm0\fdg04$. It will be part of the model for the following steps when using the radio template, and considered as a source of systematic uncertainties in the evaluation of the spectrum of Vela X. Understanding the nature of the excess at the position of PS~J0830.4$-$4449 and its relationship to Vela X is beyond the scope of this paper and left for future work.

\subsection{Full Energy Range Analysis and Spectra}\label{sec:spectrum}

We combine the best-fit models from the analyses in the LE and HE ranges and we fit them to the LAT data over the whole energy range from 10~GeV to 2~TeV. For Vela X we include two morphological components: the radio template and the disk fit to the LAT data in the HE range. For both we model the spectrum as a power law. \highlight{We tested both components for spectral curvature using as model a power law with exponential cutoff, but this resulted in an insignificant likelihood improvement.}. Other sources in the region are taken from the LE model, or from the HE model if not present in the LE model. Newly-found source candidates are considered to be the same source (thus, taken from the LE model) if their positions are within 0\fdg2. We eliminate newly-found source candidates with $\mathrm{TS}<9$ over the full energy range. This leaves in the final model only one newly found source candidate, \highlight{with $\mathrm{TS}=9.3$} { (PS~J0842.7$-$4443, see Table~\ref{tab:candidate-sources})}, plus PS~J0830.4$-$4449. \highlight{The Vela pulsar is included in the model with free spectral parameters}. We will refer to this model as model~A.

To asses the impact of the assumptions \highlight{about} the morphological representation of Vela~X on the spectral properties we consider two alternative models:
\begin{itemize}
\item in model~B we replace the radio template and PS~J0830.4$-$4449 with the disk fit to the LAT data in the LE range;
\item in model~C we replace the HE disk with the H.E.S.S. template.
\end{itemize}
From the fit of the three models to the data, we obtain the results shown in Table~\ref{tab:spectrum-all}.

\begin{table*}
\caption{\label{tab:spectrum-all}Results from the Spectral Analysis of Vela~X in the Full Energy Range $>10$~GeV: properties of the LE and HE components for different morphological models.}
\centering
\begin{tabular}{llccc}
\hline\hline
							&								& Model A		& Model B		& Model C \\
\hline
$\Delta \log \mathcal{L}$ (d.o.f.)	&								& 0	(0)			& $-13.5$ (-1)		& $-16.2$ (-3) \\
\hline
LE component				& Energy Flux\tablefootmark{a}	& $2.0\pm0.5$	& $2.5\pm0.5$ 	& $2.6\pm0.4$ \\
							& Spectral Index				& $2.19\pm0.16$	& $2.14\pm0.13$	& $1.98\pm0.16$\\
                            & TS							& 74.3			& 86.0			& 100.3\\
\hline
HE component				& Energy Flux\tablefootmark{a}	& $2.6\pm0.9$	& $2.5\pm0.9$ 	& $2.1\pm1.2$ \\
							& Spectral Index				& $0.9\pm0.3$	& $1.1\pm0.3$	& $1.2\pm0.3$\\
                            & TS							& 32.9			& 34.8			& 6.4\\
\hline
\end{tabular}
\tablefoot{The models are described in Section~\ref{sec:spectrum}. Different models are compared based on the difference of the likelihood logarithm ($\Delta \log \mathcal{L}$), and the difference in number of degrees of freedom (d.o.f.) associated with Vela X, taking model A as reference. The detection significance of a component is given by $\mathrm{TS} = 2 \times (\log \mathcal{L} - \log \mathcal{L}_0)$, where $\mathcal{L}$ and $\mathcal{L}_0$ are the maximum likelihood values of the model including or not including the component, respectively.\\
\tablefootmark{a}{$10^{-5}$~MeV~cm$^{-2}$~s$^{-1}$, integrated for energies between 10~GeV and 2~TeV.}
}
\end{table*}

Model A is confirmed to be the best representation of the data also over the whole energy range, with AIC~$-25.0$ and $-26.4$ with respect to models B and C, respectively. We can also compare model A with the spatial models of the Vela X region used in the previous catalogs that cover the same energy range, 3FHL \citep{3FHL} and FGES \citep{FGES}. In order to do so we replace the radio template, the HE disk and PS~J0830.4$-$4449 with the extended disks used in the catalogs, i.e., 3FHL~J0833.1$-$4511e, and  FGES~J0830.3$-$4453 plus FGES~J0832.0$-$4549, respectively. We obtain a decrease of $\log \mathcal{L}$ of 91.2 (27.4) for 6 (1) fewer degrees of freedom for the 3FHL (FGES) model, thus AIC=~$-170.4$ (-52.8) favors model A from this work as the best representation of the Vela X region. The full energy range fit with model A results in the detection of the soft component with morphology described by the radio template with a significance of $8.4\sigma$, and of the hard component with morphology described by the HE disk (Table~\ref{tab:morphology}) with a significance of $5.4\sigma$ (see Section~\ref{sec:ebands} for details about the conversion from TS to significance). 

We also evaluate systematic errors on the spectral results due to the LAT effective area uncertainties by applying the \emph{bracketing IRFs} method \citep{LATperf2012}. For the dataset we use, and neglecting energy dispersion in the analysis, the effective area systematic uncertainties\footnote{\url{https://fermi.gsfc.nasa.gov/ssc/data/analysis/scitools/Aeff_Systematics.html}.} are estimated to be 5\% for energies between 10~GeV and 100 GeV, and then to increase as a function of energy $E$ as: $5\% + 10\% \times [\log_{10} (E/1\;\mathrm{MeV}) -5]$. The effective area is varied within this uncertainty band according to Equation~28 in \citet{LATperf2012}. For the flux uncertainties we use $B(E) = \pm 1$, while for the uncertainties on the spectral index we adopt the expression in Equation~29 of \citet{LATperf2012} with the decorrelation energies $E_0=24$~GeV and 640~GeV for the LE and HE components, respectively. Note that there is an additional uncertainty in the absolute energy scale amounting to $+4\%/-10\%$ \citep{LATperf2012} not accounted for in these estimates.

\begin{figure*}
   \centering
  \includegraphics[width=0.8\textwidth]{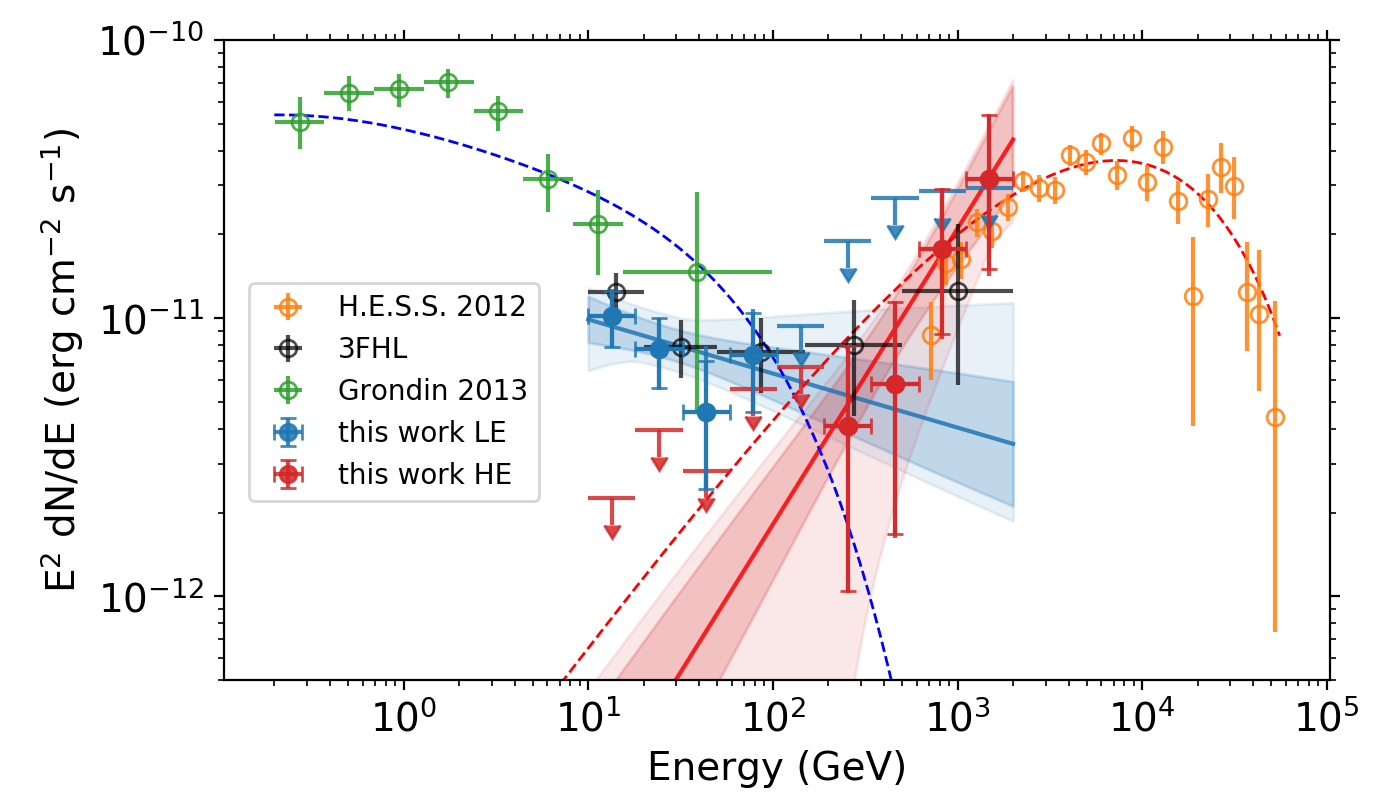}
   \caption{Spectral Energy Distribution (SED) of Vela X. We show separately the spectra for the low-energy (LE) and high-energy (HE) components as derived in this work. The lines and shaded bands show the best-fit power laws for each component with their uncertainties, based on the fit using model A. The darker shaded band corresponds to 68\% statistical uncertainties only, while the lighter shaded band corresponds to the sum in quadrature of statistical uncertainties and systematic uncertainties from the morphological representation of the sources (models B and C) and the LAT effective area (see Section~\ref{sec:spectrum} for details). The points and 95\% confidence level upper limits show the binned SED. For points capped error bars show the 68\% statistical uncertainties only, while the uncapped error bars show the sum in quadrature of statistical uncertainties and systematic uncertainties. Upper limits include systematic uncertainties as well. We also show the overall SEDs of Vela~X from H.E.S.S. \citep{hessvelax2012}, from the LAT measurements $< 100$~GeV \citep{grondin2013}, and from the 3FHL catalog \citep{3FHL}. The dashed lines \highlight{show} the predictions of the radiative model described in Section~\ref{sec:discuss}.
   }
              \label{fig:spec}%
    \end{figure*}

A summary of the spectral parameters including systematic uncertainties is given in Table~\ref{tab:spectrum-summary}. For the following we will combine in quadrature systematic uncertainties originating from the morphological representation of Vela X and from the LAT effective area.
\begin{table*}
\caption{\label{tab:spectrum-summary}Results from the Spectral Analysis of Vela~X in the Full Energy Range $>10$~GeV: summary of the properties of the LE and HE components including systematic uncertainties.}
\centering
\begin{tabular}{llcccc}
\hline\hline
Component					& Parameter						& Value			& \multicolumn{3}{c}{Uncertainties} \\
							&								&				& Statistical 	& Morphology 		& Effective Area \\
\hline
LE component				& Energy Flux\tablefootmark{a}	& 2.0			& 0.5			& $+0.6/-0$	& 0.2\\
							& Spectral Index				& 2.19			& 0.16			& $+0/-0.21$	& 0.05\\
\hline
HE component				& Energy Flux\tablefootmark{a}	& 2.6			& 0.9 			& $+0/-0.4$	& $+0.4/-0.3$\\
							& Spectral Index				& 0.9			& 0.3 			& $+0.3/-0$	& 0.1 \\ 
\hline
\end{tabular}
\tablefoot{Parameter values and statistical uncertainties correspond to model A from Table~\ref{tab:spectrum-all}. Systematic uncertainties due to assumptions on the morphology are evaluated from the maximum spread of parameters values with respect to models B and C in Table~\ref{tab:spectrum-all}. We also report systematic errors due to the LAT effective area uncertainties (estimated using the \emph{bracketing IRFs}, see Section~\ref{sec:spectrum} for details).\\
\tablefootmark{a}{$10^{-5}$~MeV~cm$^{-2}$~s$^{-1}$, integrated for energies between 10~GeV and 2~TeV.}
}
\end{table*}
Figure~\ref{fig:spec} shows the spectral energy distribution of the two extended components in Vela X both from the fit over the whole energy range and from a bin-by-bin fit over narrower energy intervals.
The spectrum of the HE component is significantly harder than that of the LE component (Table~\ref{tab:spectrum-summary}, Figure~\ref{fig:spec}), \highlight{consistent} also with the results obtained in the two separate energy bands in Section~\ref{sec:ebands}. Figure~\ref{fig:spec} illustrates that the sum of the fluxes from the two components matches well the overall fluxes derived in the 3FHL catalog \citep{3FHL}.

Figure~\ref{fig:spec} shows that the SED of the LE component overlaps with the SED derived from LAT data at energies $< 100$~GeV by \citet{grondin2013}. The latter study had pointed out a marginal spectral difference between the Northern and Southern part of the extended nebula in GeV gamma rays. Our statistically-limited dataset at energies $>10$~GeV does not permit \highlight{us} to address this point. The value of the spectral index of the LE component above 10~GeV, $2.19\pm0.16^{+0.05}_{-0.22}$, is fully consistent with the global value in \citet{grondin2013} obtained using a simple power law model, $2.24\pm 0.04$, but is marginally harder than the high-energy value for a broken power law model with a break at $\sim$2~GeV, $2.89 \pm 0.23 \pm 0.05$.

Figure~\ref{fig:spec} also shows that the SED of the HE disk connects to the SED measured at energies $>750$~GeV with H.E.S.S. \citep{hessvelax2012}. The spectral index of the HE component of $0.9\pm0.3^{+0.3}_{-0.1}$ is consistent with that derived from H.E.S.S. data above 750~GeV of $1.32 \pm 0.06 \pm 0.12$ within the uncertainties. 

\section{Discussion and conclusions}\label{sec:discuss}
The analysis of $\sim$9.5~years of \F~LAT data has enabled us to disentangle the two morphological/spectral components of the Vela~X PWN in the 10 GeV-2 TeV energy range, bridging the gap between previous results from the LAT \citep{grondin2013} and measurements with H.E.S.S. \citep{hessvelax2012}. 

At low energies, a soft component (spectral index $2.19\pm0.16^{+0.05}_{-0.22}$) extends over a region of radius $1\fdg36\pm0\fdg04$, consistent with the extension of the radio/microwave nebula. At high energies, a component with a very hard spectrum (spectral index $0.9\pm0.3^{+0.3}_{-0.1}$) is concentrated over an area of radius $0\fdg63\pm0\fdg03$, that overlaps the X-ray cocoon, already established to account for the bulk of the emission at TeV energies, but with a shift of the emission centroid. Our measurements show that the latter component becomes dominant in the 100-400~GeV energy range.

In Figure~\ref{fig:spec} we show { for illustration purposes} a simple radiative model that approximately reproduces the spectra of the two components. We assume that there are two populations of relativistic electrons, each with a spectrum described by a power law with exponential cutoff
\begin{equation}
\frac{\mathrm{d}N}{\mathrm{d}E} = N_0 \left(\frac{E}{E_0}\right)^{-p} \exp\left[-\left(\frac{E}{E_\mathrm{cut}}\right)^\alpha\right].
\end{equation}
The relativistic electrons produce IC emission upscattering photons from the Cosmic Microwave Background (CMB), far infrared emission from dust, and starlight. For the two latter components we assume greybody spectra with temperatures of 30~K and 3000 K, respectively, and energy densities of 0.2 eV~cm$^{-3}$ and 0.3 eV~cm$^{-3}$, respectively. We use this parametrization to approximate the local interstellar radiation field \citep[e.g.,][]{porter2008,popescu2017}. \highlight{As shown, e.g., in \citet{grondin2013}, dust and star emission has a sizable impact on the IC emission from the GeV nebula. Therefore, detailed modeling of the gamma-ray emission from Vela~X would require one to take the uncertainties in the radiation fields into account. This is, however, beyond the scope of this paper { and left for future work}.} We calculate IC emission using the formulae by \citet{khangulyan2014} as implemented in the \texttt{naima} Python package by \citet{zabalza2015}.

For the extended nebula/LE component we set $p=1.7$, as required by radio observations \citep[e.g.,][]{hinton2011,grondin2013}. In order to reproduce the LAT spectrum including our new measurements we assume $E_\mathrm{cut}=30$~GeV and $\alpha=0.6$. Our results confirm the roll-over of the electron spectrum at a few tens of GeV found in previous studies. As discussed in \citet{hinton2011}, a steepening of the spectrum at these energies is difficult to explain with energy losses due to synchrotron radiation or IC scattering given the physical conditions in the nebula, and there is no reason to expect it from the particle acceleration mechanism. Additionally, the slower-than-exponential cutoff is not consistent with radiative cooling. Therefore, the roll-over of the spectrum is more likely the result of a different process, such as diffusive escape from the nebula as advocated in \citet{hinton2011}.

For the X-ray cocoon/HE component we take $p=1.2$, $E_\mathrm{cut}=30$~TeV, and $\alpha=1$. These values are consistent with previous results based on X-ray and TeV \highlight{gamma-ray} measurements \citep[e.g.,][]{tibaldo2017}. { However, previous measurements were mostly constraining multi-TeV parent electron energies, while here we extend the spectral measurement to lower energies, well below the high-energy spectral cutoff}. For a power-law electron spectrum, expected from the most common models of particle acceleration in PWNe \citep[e.g.,][]{sironi2017}, the very hard spectrum of the gamma-ray emission is indicative of electrons that have not experienced yet significant cooling due to radiative losses. This is expected given the age of the cocoon of $\sim$10~kyr \citep{blondin2001}, since, for the magnetic field strength of 5~$\upmu$G \citep{dejager2008}, the cooling time of electrons with energies $<10$~TeV is $>30$~kyr.

For the same hypothesis of a power-law electron spectrum, in the \highlight{Thomson} regime the IC spectral index $\Gamma$ is related to the electron spectral index $p$ as: $\Gamma = (1+p)/2$. The electron spectral index $p \simeq 2.2-2.4$ predicted at late ($>10$~kyr) times for Fermi acceleration at relativistic shocks \citep{achterberg2001,keshet2005,sironi2017} would correspond to an IC spectral index $\Gamma \simeq 1.6-1.7$. The value we find of $0.9\pm0.3^{+0.3}_{-0.1}$ hints that the electron spectrum may be harder in the Vela~X cocoon.

Several mechanisms discussed in the literature could produce such a harder spectrum. \citet{bykov2017} predict hard spectra in Vela~X for electrons reaccelerated  in the converging flow system formed by the plasma outflowing from the wind termination shock and the plasma inflowing from the bowshock caused by the motion of the pulsar within its parent SNR. Another mechanism, often invoked to explain the hard TeV spectra of active galactic nuclei, but possibly applicable also to PWNe, is stochastic particle acceleration combined with strong synchrotron and IC energy losses that produce a Maxwellian-like electron spectrum \citep{schlickeiser1985,stawarz2008}. Furthermore, much attention has been given lately to the role played by particle acceleration in magnetic reconnection in PWNe. Magnetic reconnection models generally predict spectra harder than shock acceleration for high wind magnetizations $\sigma\gtrsim 10$ \citep{sironi2014}. This condition may not be realized in Vela~X, e.g., \citet{buehler2016} estimate $\sigma \simeq 3$. However, \citet{guo2014} argue that hard power-law spectra may also be achieved with $\sigma \sim 1$ in sufficiently large reconnection layers. Finally, \citet{horns2006} proposed a model for TeV emission from Vela~X dominated by relativistic ion inelastic collisions. Their model predicts a very hard spectrum \highlight{at a few hundred} GeV. Moreover, the coexistence of ion and IC emission with different spectra may help to explain the shift of the emission centroid found between the LAT and H.E.S.S. measurements.

Spatial and spectral results from this analysis can be incorporated in future multiwavelength studies of Vela~X, and compared to comprehensive physical models of the nebula in order to further constrain the mechanisms of particle acceleration and transport. The spatial and spectral properties of Vela~X in the energy range above a few tens of GeV are soon expected to be measured in greater detail by the Cherenkov Telescope Array \citep[CTA,][]{CTA2011}.

\begin{acknowledgements}
The \textit{Fermi} LAT Collaboration acknowledges generous ongoing support
from a number of agencies and institutes that have supported both the
development and the operation of the LAT as well as scientific data analysis.
These include the National Aeronautics and Space Administration and the
Department of Energy in the United States, the Commissariat \`a l'Energie Atomique
and the Centre National de la Recherche Scientifique / Institut National de Physique
Nucl\'eaire et de Physique des Particules in France, the Agenzia Spaziale Italiana
and the Istituto Nazionale di Fisica Nucleare in Italy, the Ministry of Education,
Culture, Sports, Science and Technology (MEXT), High Energy Accelerator Research
Organization (KEK) and Japan Aerospace Exploration Agency (JAXA) in Japan, and
the K.~A.~Wallenberg Foundation, the Swedish Research Council and the
Swedish National Space Board in Sweden.
 
Additional support for science analysis during the operations phase is gratefully
acknowledged from the Istituto Nazionale di Astrofisica in Italy and the Centre
National d'\'Etudes Spatiales in France. This work performed in part under DOE
Contract DE-AC02-76SF00515.\newline

GF acknowledges the support that made his participation to this project possible: funding from the Erasmus Traineeship program, hospitality by the Max-Planck-Institut für Kernphysik, and mentorship by R. Bonino at the University of Torino. RZ acknowledges the Alexander von Humboldt Foundation for financial support, and the Max-Planck-Institut für Kernphysik as hosting institution.\newline

This research made use of several community-developed Python packages: Astropy \citep{astropy2013}, Fermipy \citep{wood2017}, naima \citep{zabalza2015}, scikit-image \citep{scikit-image2014}, aplpy \citep{aplpy2012}, matplotlib \citep{matplotlib}. This research also employed the public datasets from the \textit{ROSAT} All-Sky Survey (\url{http://www.mpe.mpg.de/xray/wave/rosat/}) and the \textit{Planck} mission (\url{http://www.esa.int/Planck}).\newline

The authors thank N. Bucciantini and B. Cerutti for insightful discussions on the interpretation of the results.
\end{acknowledgements}

\bibliographystyle{aa}

\appendix

\section{Source candidates in the region around Vela X}

\begin{table}
\caption{\label{tab:candidate-sources}. { Coordinates of candidate sources outside Vela~X found during the morphological analysis in Section~\ref{sec:ebands}.}}
\centering
{
\begin{tabular}{lccc}
\hline\hline
Name				& R.A.				& Dec.				& TS\\
\hline
PS J0821.5$-$4658	& $125.40\pm 0.07$	& $-46.98\pm 0.09$	& 17.2\\
PS J0823.3$-$4247	& $125.83\pm 0.08$	& $-42.79\pm 0.07$	& 34.0\\
PS~J0839.5$-$4343	& $129.88\pm 0.16$	& $-43.73\pm 0.13$	& 13.4\\
PS~J0842.7$-$4443\tablefootmark{$\dagger$} 	& $130.69\pm 0.09$	& $-44.73\pm 0.09$	& 11.6\\
PS~J0847.7$-$4559	& $131.95\pm 0.08$	& $-45.99\pm 0.08$	& 15.2\\
\hline
PS~J0823.6$-$4328	& $125.92\pm 0.09$	& $-43.48\pm 0.07$  & 11.5 \\
\hline
\end{tabular}
}
\tablefoot{{ The upper section refers to the analysis in the LE range, the lower section to the HE range.\\
\tablefootmark{$\dagger$}{Source included also in the full-energy range analysis.}}
}
\end{table}

\end{document}